\documentclass[
aip,apl,reprint,
amsmath,amssymb
]{revtex4-1}

\usepackage{graphicx}
\usepackage{siunitx}

\begin{document}

\title{MEMS approach to low-frequency broadband acoustic metamaterials}

\author{Lixiang Wu}
  \email{wulx@hdu.edu.cn}
  \altaffiliation[Present address: ]{Department of Biomedical Engineering, University of Southern California, Los Angeles, CA 90089, USA}
\author{Quansheng Sun}
\affiliation{School of Electronics and Information, Hangzhou Dianzi University, \newline Hangzhou 310018, China}

\author{Qifa Zhou}
\affiliation{Department of Biomedical Engineering, University of Southern California, \newline Los Angeles, CA 90089, USA}

\author{Gaofeng Wang}
  \email{gaofeng@hdu.edu.cn}
\affiliation{School of Electronics and Information, Hangzhou Dianzi University, \newline Hangzhou 310018, China}

\date{\today}

\begin{abstract}
We experimentally demonstrate a simple micro-electro-mechnical systems (MEMS) approach to acoustic metamaterials and have observed the average increase of nearly 6 dB beyond the classic law for sound attenuation at low frequencies from 200~Hz to 1200~Hz.
Here, we have also found that the MEMS metastructure, especially the cavity-backed micromembrane, contributes to 22.3\% gain of sound transmission loss (STL) with a fill factor of perforation of less than 7.6\%.
\end{abstract}

\maketitle

Acoustic metamaterials with locally resonant characterization lead to narrow-band response naturally.~\cite{liu2000locally,fang2006ultrasonic}
The STL peaks at the antiresonance frequency between the first two resonances, which is caused by the frequency dispersion of effective mass density and/or bulk modulus.
To obtain broadband performance on sound adsorption or attenuation, STL peaks need to be combined, for example, local resonators of different resonances are assembled to exhibit a flat absorption spectrum of high STL at low frequencies.~\cite{yang2017optimal}
For membrane-type acoustic metamaterials,~\cite{yang2008membrane} however, such broadband sound attenuation can also be possibly attributed to stiffness-controlled effects on vibrating membranes at frequencies below the first natural freuqency, especially for those small membranes without attached proof mass.~\cite{sui2015lightweight,peiffer2015comment}
Because, theoritically, a no-mass-attached micromembrane clamped rigidly can introduce high STL and excellent acoustical performance at low frequencies with minimum weight penalty, when the size of no-mass-attached membranes is small enough to extend the stiffness-controlled region to cover a much wider range of low frequencies.~\cite{li2018enhanced}

To fabricate micromembranes with diameter of sub-millimeters, the MEMS technology is preferred.
In microfabrication, the release of micromembranes by the etch process would leave or produce microcavities.
As fundamental microstructures for most of MEMS sensors, both micromembranes and microcavities can help to improve the performance of sound attenuation.
Micromembranes operating in stiffness mode induce high STL meanwhile microcavities are likely to increase STL due to significant thermoviscous effects, which are quite common in MEMS.
In particular, air damping such as squeeze-film damping can provide considerable damping and elastic forces and cause strong energy dissipation.~\cite{bao2007squeeze}
Recently, it has been reported that the thermoviscous dissipation can effectively reduce sound transmission through a metasurface of hybrid resonance~\cite{jiang2017thermoviscous} and the thermoviscous losses arising from a waveguided air gap can induce stronger absorption when the thickness of air gap decreases to tens to hundreds of microns.~\cite{starkey2017thin}

Compared to acoustic membranes with diameter of centimeters of the state-of-the-art membrane-type metamaterials, MEMS micromembranes with diameter of millimeters or sub-millimeters can be compatible with integrated circuits~\cite{fischer2015integrating} and have the potential applications in microelectrical packaging and MEMS security.
Thus, we explore a MEMS approach to low-frequency broadband acoustic metamaterials and here report a simple experiamental demonstration of MEMS metastructures for increasing STL at low frequencies from 200~Hz to 1200~Hz.

\begin{figure}[!b]
\centering\includegraphics[width=.9\columnwidth]{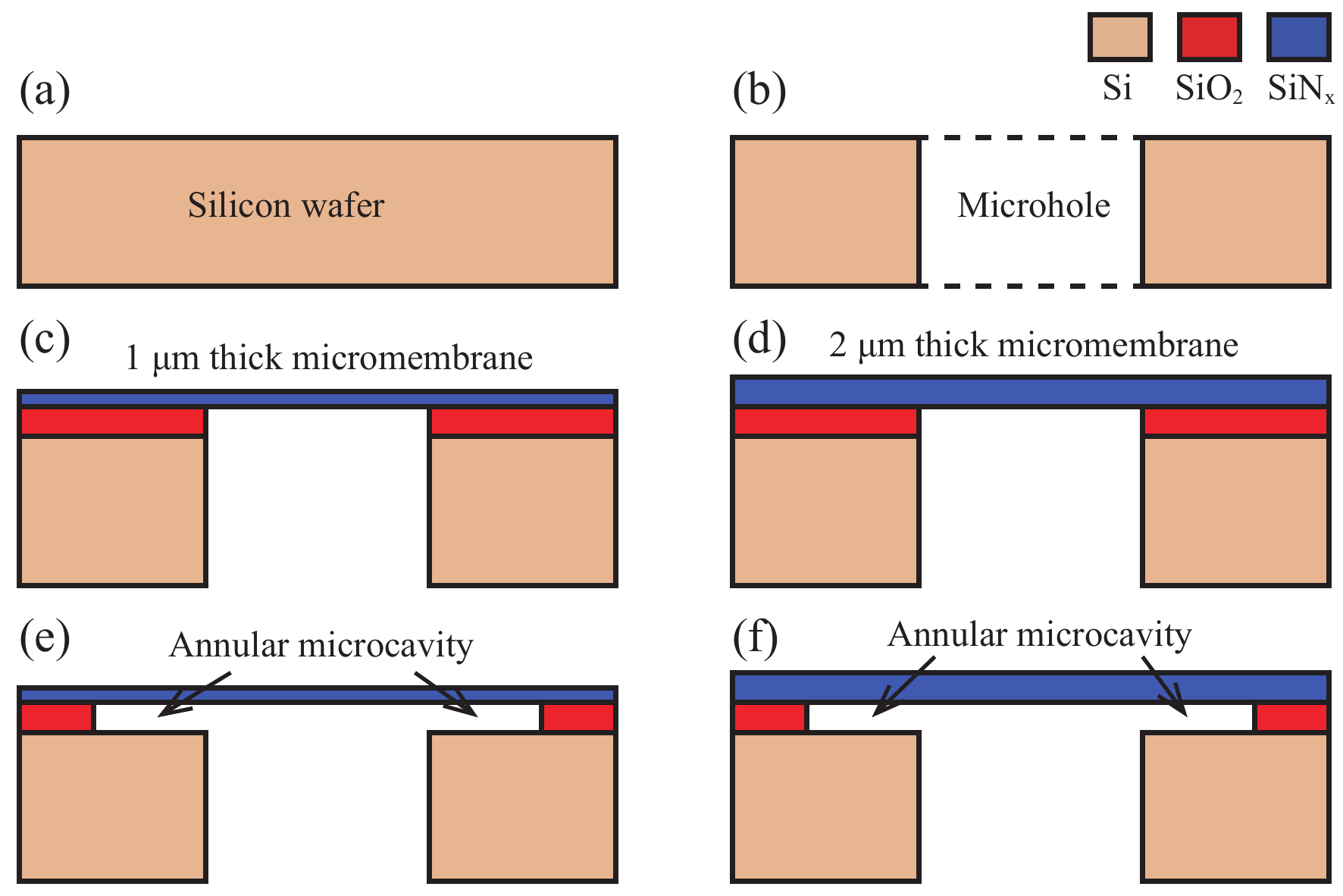}
\caption{\label{fig:metastructure} MEMS metastructures: (a) raw wafer, (b) perforated wafer, (c-d) perforated wafers capped by \SI{1}{\micro\metre} and \SI{2}{\micro\metre} thick micro-membranes, (e-f) perforated wafers capped by \SI{1}{\micro\metre} and \SI{2}{\micro\metre} thick micromembranes backed by microcavities.}
\end{figure}

\begin{figure}[!tb]
\centering\includegraphics[width=.95\columnwidth]{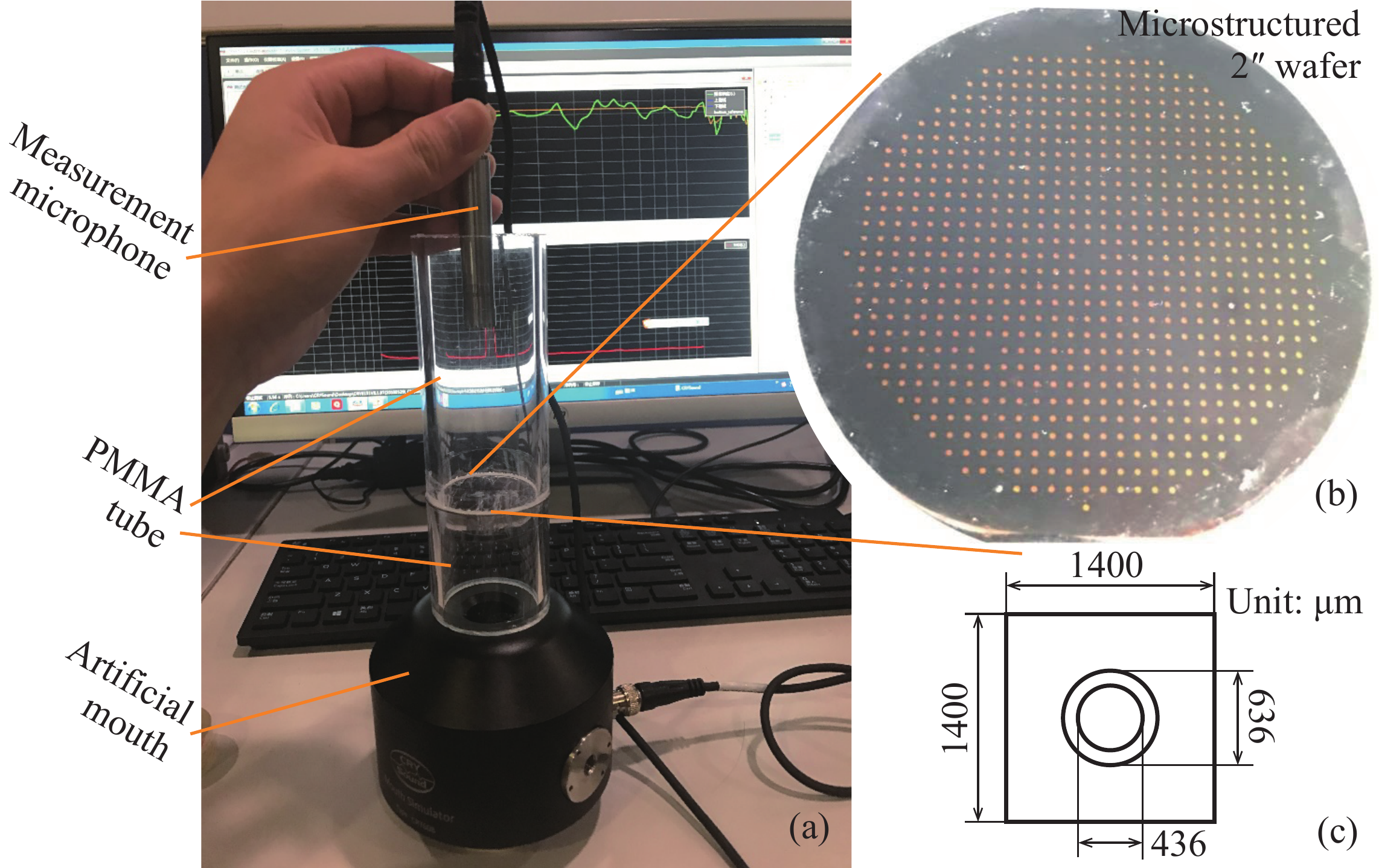}
\caption{\label{fig:expinstall} (a) Simple experimental installation for sound attenuation measurement, (b) pattern of microstructured 2'' wafer, and (c) dimensions of a unit cell.}
\end{figure}

In the experiment, six groups of microstructured silicon wafers (2 inch, 300~μm thick) were prepared to study the sound attenuation performance of MEMS metastructures including microholes, micromembranes, and microcavities, which are illustrated in Figure~\ref{fig:metastructure}.
And all the MEMS metastructures were patterned in array on a 2 inch silicon wafer, where the pitch is 1.4~mm$\times$1.4~mm and the fill factor of microholes at the silicon wafer is less than 7.6\% (see Figure~\ref{fig:expinstall}(b-c)).
The microholes with diameter of \SI{436}{\micro\metre} were perforated by the deep reactive ion etch (DRIE) technology.
To stop sound transimission via air flow through microholes, micromembranes with thickness of \SI{1}{\micro\metre} or \SI{2}{\micro\metre} were capped on the top of those microholes (see Figure~\ref{fig:metastructure} (c-d)) by chemical vapor deposition of thin films (SiN$_x$/SiO$_2$).
Moreover, the \SI{2}{\micro\metre} thick annular microcavity with width of about \SI{100}{\micro\metre} is etched by buffered hydrofluoric acid beneath the micromembrane to introduce air damping (see Figure~\ref{fig:metastructure}(e-f)).

A simple experimental installation is illustrated in Figure~\ref{fig:expinstall}(a) and used to measure sound transmission through a microstructured wafer, which consists of two \textit{Poly(methyl methacrylate)} (PMMA) tubes (2 inch, 5~mm thick), an artificial mouth and a measurement microphone.
The to-be-tested wafer is clamped between two PMMA tubes and the sound radiated from the bottom artificial mouth transmits through the whole installation then arrives at the measurement microphone suspended in the top tube.
The tube is not a perfect waveguide tube and the silicon wafer is not an absolute rigid plate, therefore very careful comparisons would be conducted among the six groups to evaluate the real contribution of MEMS metastructures to sound attenuation.

\begin{figure}[!tb]
\centering\includegraphics[width=.75\columnwidth]{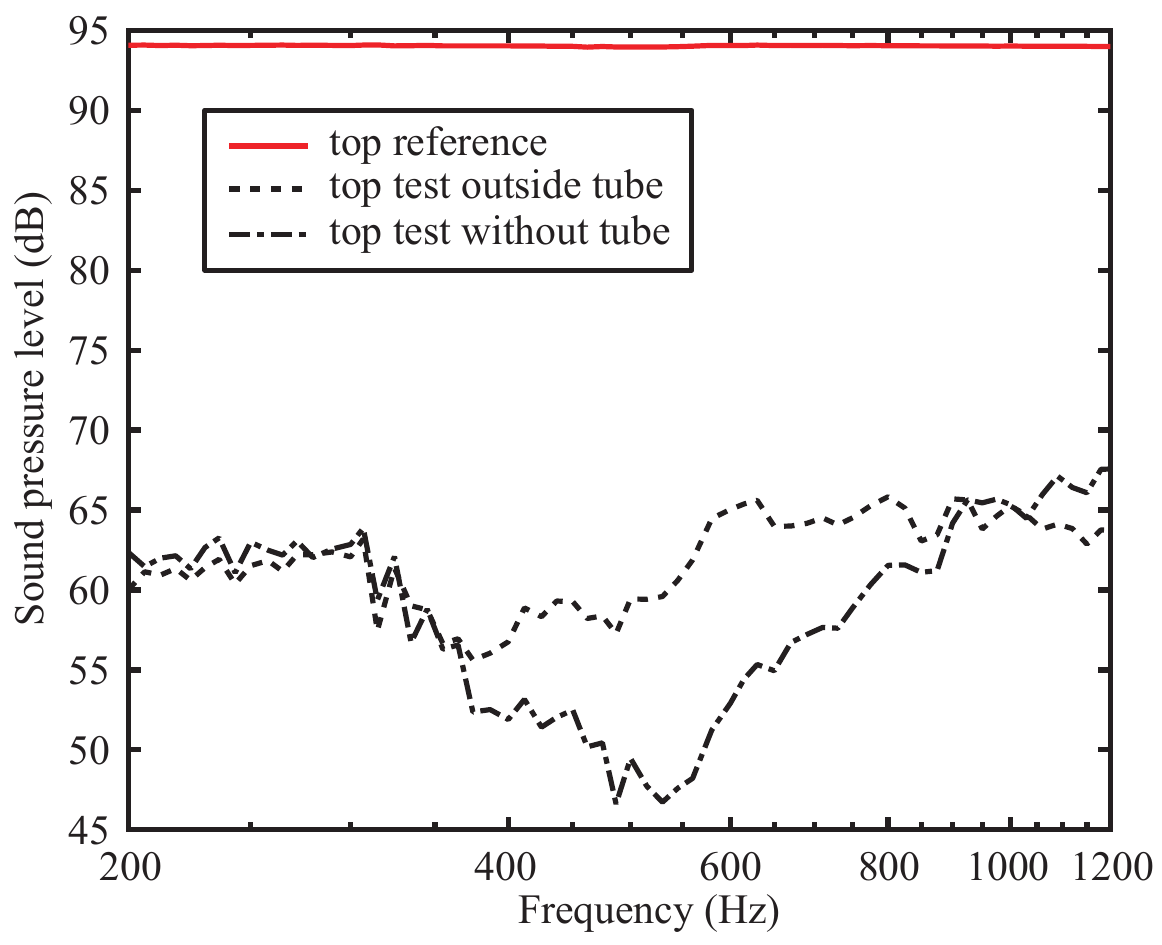}
\caption{\label{fig:spl} Reference measurement for sound transmission through the PMMA tube from the bottom artificial mouth to the top side.}
\end{figure}

First, references are set up for sound transmission through the PMMA tube from the bottom artificial mouth to the top side and the measurement result is shown in Figure~\ref{fig:spl}, where the top reference is measured with only the PMMA tube, the top reference outside tube is measured outside the tube, and the top reference without tube is measured in the free field.
The result indicates that the PMMA tube has a sound insulation performance of more than 40~dB at the range of 200--1200~Hz and the sound transmitted in the tube is waveguided.

\begin{figure}[!tb]
\centering\includegraphics[width=.75\columnwidth]{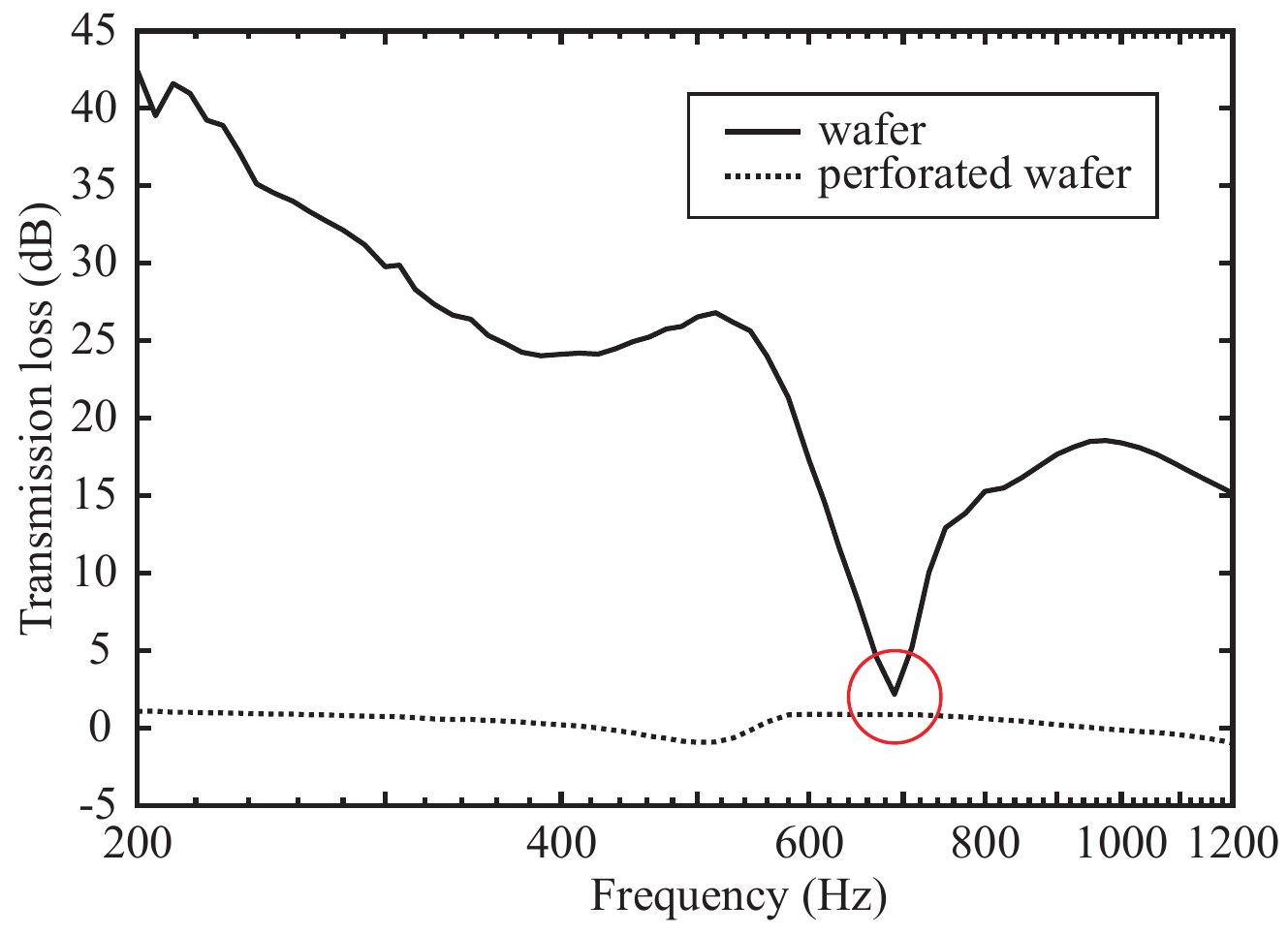}
\caption{\label{fig:stl} Sound transmission losses of the 2'' silicon wafer and the perforated one.}
\end{figure}

\begin{figure}[!b]
\centering\includegraphics[width=.9\columnwidth]{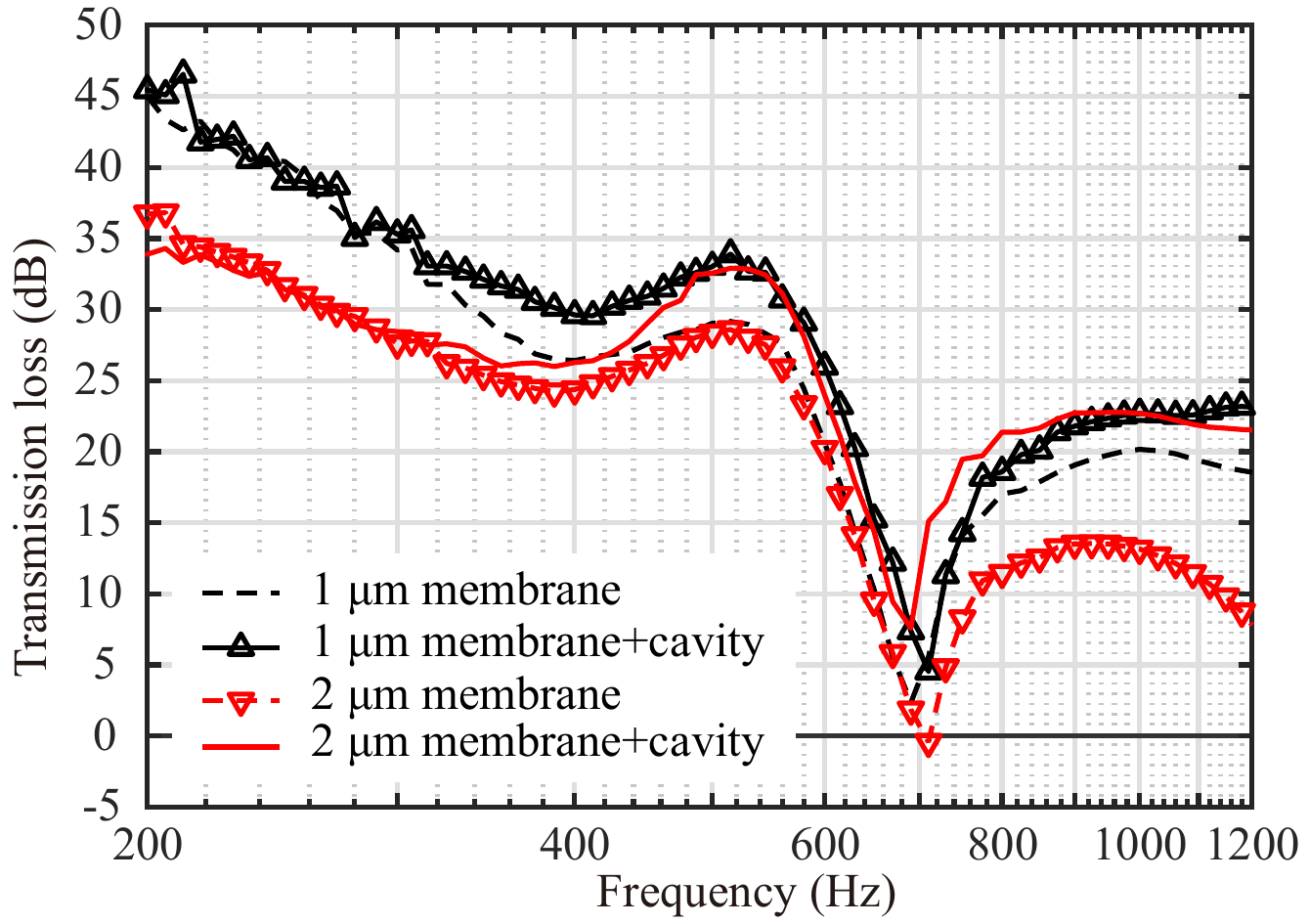}
\caption{\label{fig:stl2} Sound transmission losses of four MEMS metastructures.}
\end{figure}

Then, STLs of the silicon wafer and the perforated wafer are measured as is shown in Figure~\ref{fig:stl}.
The wafer has a high STL except at the dip around 700~Hz, while the perforated wafer is almost transparent with a negligible impedance.
The STL dip is likely to occur at the first resonant frequency of the to-be-tested wafer~\cite{zhou2010boundary} and an average STL of about 30~dB is measured below the first resonant frequency (\textit{i.e.}, 200--700~Hz).
Compared with the above two wafers, other four microstructured wafers have micro-membranes with/without micro-cavities and show exotic behaviours on sound insulation (see Figure~\ref{fig:stl2}).
Overall, STL curves for cavity-backed membrane of \SI{1}{\micro\metre} thickness and \SI{2}{\micro\metre} thick membrane roughly contour the upper and lower limits.
At the vicinity of frequency dip, the STL curve of the \SI{2}{\micro\metre} thick membrane falls down and crosses the zero point, indicating that the sound transmission is significantly enhanced to be even better than the top reference; the \SI{2}{\micro\metre} thick cavity-backed membrane has the highest STL among the four cases shown in Figure~\ref{fig:stl2}.

\begin{figure}[!b]
\centering\includegraphics[width=.75\textwidth]{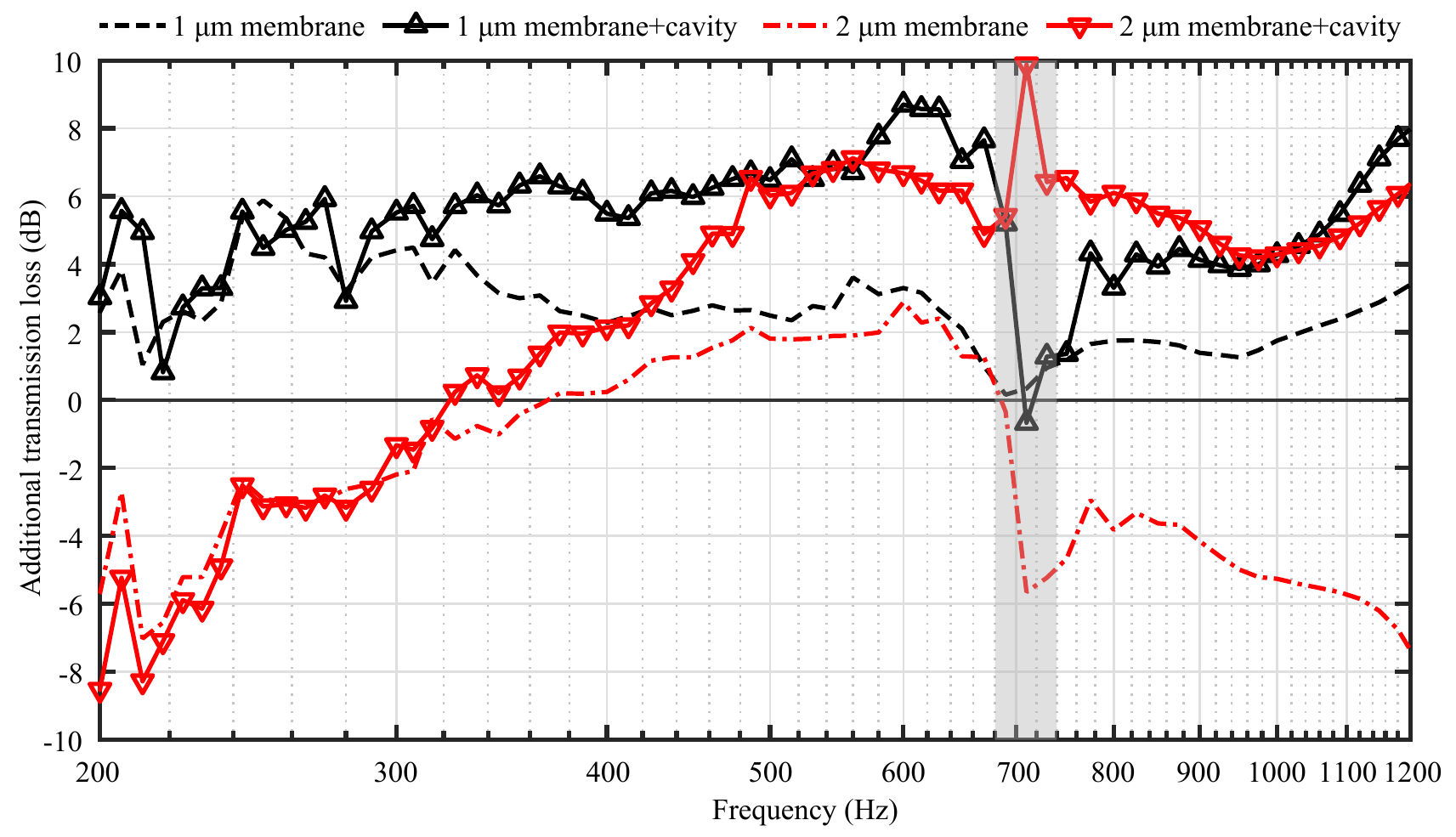}
\caption{\label{fig:astl} Additional sound transmission losses of four MEMS metastructures compared to that of the raw wafer.}
\end{figure}

Furthermore, the contribution of MEMS metastructures to sound attenuation is defined as the additional STL, which is the STL gain compared to the measurement result of the raw wafer.
Figure~\ref{fig:astl} displays the curves of additional STL in correspondence to the metastructures shown in Figure~\ref{fig:metastructure} (c-f).
As is shown in Figure~\ref{fig:astl}, the frequencies can be roughly divided into three regions, \textit{i.e.}, stiffness-controlled region (200--680~Hz), resonance-controlled region (680--740~Hz), and mass-controlled region (740--1200~Hz).
In the stiffness-controlled region, the \SI{1}{\micro\metre} thick SiN$_x$ membrane with annular microcavity performs better on sound insulation than the \SI{2}{\micro\metre} thick membrane backed by microcavity; and the former has a slightly increasing curve with an average additional STL of nearly 6~dB while the latter soars up from -6~dB to 6~dB as the frequency increases.
The gap of additional STL between the cavity-backed membranes and the no-cavity-backed ones starts to get wider from 300 Hz since the damping effects become stronger with increase of the sound frequency.
In the resonance-controlled region, the additional STL curves dip suddenly except for the curve of the \SI{2}{\micro\metre} thick cavity-backed membrane that peaks up to 10~dB.
Even more exceptional, the additional STL of the \SI{2}{\micro\metre} membrane falls down to almost -6~dB and this trend extends to the mass-controlled region.
In the mass-controlled region, the additional STLs of the cavity-backed micromembranes are 5~dB higher than that of a raw wafer, implying that it produces 5~dB gain beyond the classic mass law.
The merge of STL curves of the cavity-backed membranes of \SI{1}{\micro\metre} and \SI{2}{\micro\metre} could give rise to an average 6~dB gain of STL in the frequency range from 200~Hz to 1200~Hz, across all three regions.

\onecolumngrid 

\twocolumngrid

In this work, the STL of a specific microstructured wafer consists of the STL of the wafer with diameter of 50~mm and the additional STL of the specific MEMS metastructure in the sub-millimeter scale.
The primary contribution to STL is the stiffness-controlled energy dissipation of the wafer, whereas, the MEMS metastructure contributes to 22.3\% gain of STL with the fill factor of perforation of less than 7.6\%.
Moreover, there should be a trade-off between the stiffness of wafer and the fill factor of perforation, since the larger fill factor would induce the lower stiffness and lighter weight.
We conclude that the optimization with regard to STL could be conducted on multiscale metastructures.
Herein simply demonstrated is two-scaled.
The multiscale combination of energy dissipation would benefit from the scaling effects of physical model.
For example, thermoviscous effects become prominent only at ultra-subwavelength or usually microscale.

\begin{acknowledgments}
L. Wu thanks Gengxi Lu for discussions on acoustic metamaterials and acknowledges support from the program of China Scholarship Council (No. 201708330397).
G. Wang acknowledges support from the National Natural Science Foundation of China (No. 61411136003).
\end{acknowledgments}

\providecommand{\noopsort}[1]{}\providecommand{\singleletter}[1]{#1}%

\end{document}